\documentclass[twocolumn,prb,aps,showpacs,preprintnumbers,amsmath,amssymb ]{revtex4}

\usepackage{graphicx}
\usepackage{dcolumn}
\usepackage{bm}

\input epsf

\newcommand{\refeq}[1]{(\ref{#1})}

\newcommand{\dif}{{\mathrm  d}}

\newcommand{\bnabla}{\bm\nabla}

\newcommand{\bE}{{\bf E}}

\newcommand{\bI}{{\bf I}}

\newcommand{\bT}{{\bf T}}

\newcommand{\br}{{\bf r}}

\newcommand{\bn}{{\bf n}}
\newcommand{\bt}{{\bf t}}

\newcommand{\bu}{{\bf u}}
\newcommand{\bv}{{\bf v}}

\newcommand{\rhat}{{\bf \hat r}}

\newcommand{\trac}{\bm{\tau}}
\newcommand{\tracT}{\Theta}
\newcommand{\bS}{{\bf y}}

\newcommand{\im}{{\mathrm i}}

\newcommand{\eq}{{\mathrm{eq}}}

\newcommand{\brac}[1]{\left(#1\right)}

\newcommand{\visrat}{{\lambda}}

\newcommand{\eps}{{\varepsilon}}

\newcommand{\mem}{{\mathrm{mem}}}

\newcommand{\out}{{\mathrm{out}}}
\newcommand{\ins}{{\mathrm{in}}}

\newcommand{\deform}{{\mathrm{\dot\gamma}}}

\newcommand{\rot}{{\mathrm{r}}}

\newcommand{\half} {{\frac{1}{2}}}

\begin{document}

\title{Dynamics of a viscous vesicle  in linear flows}

\author{Petia M. Vlahovska}
\altaffiliation[Present address: ]{Thayer School of Engineering, Dartmouth College, Hanover, NH 03755.}
\email{petia.vlahovska@dartmouth.edu}
\author{Ruben Serral Gracia}
\affiliation{%
Max-Planck Institute of Colloids and Interfaces, Theory Department, 14424 Potsdam-Golm, Germany
}


\begin{abstract}

An analytical theory is developed to describe the dynamics of a closed
lipid bilayer membrane (vesicle) freely suspended in a general linear
flow. Considering a nearly spherical shape, the solution to the
creeping--flow equations is obtained as a regular perturbation
expansion in
the excess area.
The analysis takes into account the membrane
fluidity, incompressibility and resistance to bending. The constraint
for a fixed total area leads to a non--linear shape evolution equation
at leading order.  As a result two regimes of vesicle behavior,
tank--treading and tumbling, are predicted depending on the viscosity
contrast between interior and exterior fluid.  Below a
critical viscosity contrast, which depends on the excess area, the
vesicle deforms into a tank--treading ellipsoid, whose orientation
angle with respect to the flow direction is independent of the
membrane bending rigidity. In the tumbling regime, the vesicle
exhibits periodic shape deformations with a frequency that increases
with the viscosity contrast.  Non-Newtonian rheology such as  normal
stresses is predicted for a dilute suspension of vesicles. The theory
is in good agreement with published experimental data for vesicle
behavior in simple shear flow.

\end{abstract}
\pacs{ 47.15.G-, 83.50.-v, 87.16Dg}

\maketitle

\section{Introduction}

The dynamics of deformable particles such as drops, capsules and cells
in flow represents a long--standing problem of interest in many
branches of science and engineering, for instance because of its
relevance to the rheology of emulsions and biological suspensions such
as blood. The problem is challenging because the shape of these
``soft'' particles is not given {\it{a priori}} but is governed by the
balance between interfacial forces, e.g. due to stretching and/or
bending of the interface, and fluid stresses. The interfacial
properties, therefore, play a crucial role in the dynamics of these
particles.

The interface between two simple fluids
is governed by the surface tension, which is isotropic, in the absence of
surfactants or heating. Surface tension acts to minimize
the interfacial area; therefore the rest shape of a drop is a
sphere.  Under shear flow, the drop deforms initially into an
ellipsoid. As the flow strength increases, the drop area also
increases and the angle between the ellipsoid major axis and flow
direction decreases from $\pi/4$. If the flow strength is sufficiently
high, and the viscosity contrast is moderate, the drop breaks
up. Drop microhydrodynamics has been studied quite extensively
\cite{Rallison:1984, Stone:1994}.

Polymerized membrane  interfaces found  in synthetic capsules  and the
red blood cell, where a lipid  bilayer is attached to a scaffolding of
cytoskeletal  proteins,   exhibit  far  more   complicated  mechanical
properties.   These  membranes  behave as  thin  viscoelastic
materials  that can develop  bending moments  similar to  thin elastic
shells.  Barthes-Biesel \cite{Barthes_Biesel:1991} has reviewed the various
constitutive laws  adopted to describe the membrane  mechanics and the
effects of interfacial properties on  the capsule behavior in a linear
flow.  Yet, most  of  the  theoretical  studies treat  the  membrane  as  a
two--dimensional viscoelastic surface with {\it{no}} bending resistance.

Bending stiffness, however, plays a crucial role in the mechanics of
biological membranes, for example the equilibrium biconcave shape of
the red blood cell can not be accounted for without including bending
resistance\cite{Canham:1970}. The main structural component of the
cell membrane is the lipid bilayer and its mechanical properties are
essential to the overall cell mechanics. The pure lipid membrane
consists of two sheets of lipid molecules. The molecular thickness
imparts bending resistance. Lipid molecules are free to move within
the monolayer, and therefore, in contrast to solid--like polymerized
membranes, the lipid bilayer membrane is fluid.  In addition, since
the lipid bilayer contains a fixed number of molecules with fixed area
per molecule (under moderate stresses),
the membrane is incompressible and the total area is constant. The
enclosed volume is also constant at given osmotic conditions. The
mechanics of lipid bilayers in concisely reviewed in Powers\cite{Tom:review}.

Equilibrium mechanical properties of vesicles made of lipid bilayer
membranes are fairly well understood \cite{Seifert:1997}, and vesicle
shapes can be generated by minimizing the membrane bending energy
subject to the constraints of constant area and enclosed volume.  In
contrast, the non--equilibrium dynamics of lipid bilayer membranes has
been studied only to a limited extent.  Experimental studies
\cite{Haas-Blom-Ende-Duits-Mellema:1997, Kantsler-Steinberg:2005,
Kantsler-Steinberg:2006, Mader:2006} of vesicle behavior in unbounded
shear flow observe that in weak flows, and when the inner and outer
fluids are the same, the vesicle deforms into a tank-treading
stationary prolate ellipsoid with an inclination angle close to $\pi/4$ with respect to
the flow direction; however, in striking contrast to
drops, when the fluid inside is more viscous than outside the vesicle
undergoes a tumbling motion.  Numerical simulations
\cite{Kraus-Seifert_etal:1996, Beaucourt_etal:2004,
Noguchi-Gompper:2004} and analytical
theories\cite{Seifert:1999,Olla:2000,Misbah:2006} of vesicle
microhydrodynamics attempt to elucidate such experimental
observations.  In their classic paper Keller and
Skalak\cite{Keller-Skalak:1982} analyzed the motion of a
tank--treading ellipsoid in unbounded shear flow. The theory
qualitatively captures features like the tumbling transition, but
their assumption of a fixed particle shape casts some doubts on the
applicability to deformable vesicles; for instance, no connection can
be made to the physical properties of real membranes such as the bending
rigidity.  The free-boundary character of the problem is taken into
account in several recent works.  Shape evolution of a fluctuating
quasi--spherical vesicle was considered by Seifert\cite{Seifert:1999}
for the case where the inner and suspending fluids are the same,
i.e. there is no viscosity contrast.  The theory predicts a
stationary, tank--treading prolate ellipsoid and no transition to
tumbling motion.  The importance of viscosity contrast for the
tumbling transition was recognized by Misbah\cite{Misbah:2006}, who
showed quantitatively that a critical value of viscosity contrast
exists which separates the tank--treading and tumbling regimes. The
critical viscosity ratio was shown to decrease with the excess area
(the difference in the areas of the vesicle and an equivalent sphere
with the same volume). He also pointed out that the area constraint
leads to non-linear leading order evolution equations for the vesicle
shape, which in the case when only ellipsoidal deformation modes are
considered are independent of the membrane bending rigidity. Earlier
work by Olla\cite{Olla:2000} derived similar results for a
viscoelastic membrane.

Cell behavior in unbounded flow is of fundamental interest. However, flows in confined geometries are much more relevant to physiology, for instance
blood flow in the microcirculation. Walls affect greatly particle
microhydrodynamics, for example red blood cells migrate away from the
blood vessel walls, an observation that dates back to Poiseuille
\cite{Sutera--Skalak:1993}. The existence of a near--wall
cell--depleted region accounts for the Fahraeus--Lindqvist effect,
which is the decrease of the apparent blood viscosity in smaller
vessels\cite{Popel:2005}.
Cell
traffic between the blood stream and tissues involves cell attachment to blood vessel walls
\cite{Orsello-Hammer:2001}; examples are leukocytes during inflammatory
response,
platelets in formation of
atherosclerotic plaques,
or tumor cells in metastasis.
In order to elucidate the fundamental features of the process, several
studies have used vesicles as the simplest artificial cell
\cite{Lorz-Simson-Nardi-Sakmann:2000, Abkarian:2002,
Abkarian:2005}. They have reported that the flow--induced deformation
of adhering vesicles gives rise to a lift force that can lead to
unbinding from the substrate. The experimental data on the lift force
and vesicle migration velocity have not yet been quantitatively
compared to theoretical studies \cite{Cantat-Misbah:1999,
Sukumaran-Seifert:2001, Seifert:1999b}.

The purpose of this paper is to simultaneously include the effects of
({\it{i}}) viscosity contrast, ({\it{ii}}) membrane incompressibility and
({\it{iii}}) bending rigidity in the analysis of vesicle dynamics in
shear flows, unbounded or in the presence of a wall, in a consistent way and thereby proceed further towards a
fully quantitative description of the experimentally observed vesicle
behavior.  As a first step, the leading order small deformation of a
nearly--spherical vesicle will be considered.  However, the developed
formalism will serve as a rigorous basis for considering higher orders
in the non--linear dynamics of a vesicle in flow.

Our study extends the results of Seifert\cite{Seifert:1999} derived
for an equiviscous vesicle.
The works by Misbah\cite{Misbah:2006} and Olla\cite{Olla:2000} are
clarified, more specifically, the calculations are presented more 
explicitly,
some important results are corrected and
cast in a form that can be directly compared to experimental
measurements, and more physical insights are given for the vesicle
behavior in the tumbling regime. We demonstrate that the theory agrees
quantitatively with published experimental
data\cite{Kantsler-Steinberg:2005, Kantsler-Steinberg:2006,
Mader:2006}. We present new results for ({\it{i}}) the non--Newtonian
rheology of a vesicle suspension, in particular, we show how the
single vesicle solution serves to calculate the effective stress of a
collection of many vesicles and we predict the existence of normal
stresses, and ({\it{ii}}) vesicle migration velocity in wall-bounded
flow in the case of large distances from the boundary.

\section{Problem formulation}
Let us consider a neutrally--buoyant vesicle formed by a closed lipid
bilayer membrane with total area $A$. The vesicle is suspended in a
fluid of viscosity $\eta$ and filled with a fluid of viscosity
$\left(\visrat-1 \right) \eta$ .  Both interior and exterior fluids
are incompressible and Newtonian. The vesicle has a characteristic
size $a$ defined by the radius of a sphere of the same volume.  The
equilibrium shape of the vesicle is characterized by a small excess
area
\begin{equation}
\label{excess area}
\Delta=A/a^2-4 \pi \,.
\end{equation}
The coordinate system employed is spherical $(r,\, \theta,\, \phi)$,
with the origin coinciding with the center of mass of the vesicle.
The interface is specified by a shape function $F\brac{\br,t}$, of the
position $\br$ and time $t$, which represents the interface as the set
of points, where $F\brac{\br,t}\equiv 0$.  It is given by the relation
\begin{equation}
\label{F}
F\brac{\br,t}=r-r_{\mathrm{s}}\left(\theta, \phi, t\right)\,.
\end{equation}

\subsection{Hydrodynamics}  
\begin{figure}
\epsfxsize=2in
\epsffile{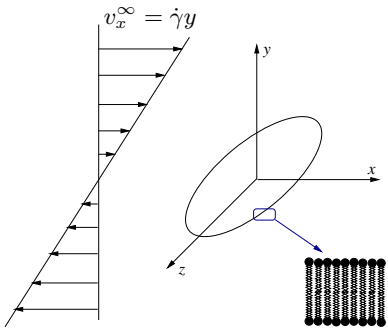}
\begin{picture}(0,0)(0,0)
\put(-110,115){$v^{\infty}_x=\dot \gamma y$}
\end{picture}
\caption{\footnotesize A sketch of a vesicle is a simple shear flow. The zoomed region of the interface illustrates the bilayer lipid structure of the membrane.}
\label{sketch}
\end{figure}

The vesicle is placed in a steady two-dimensional linear flow
\begin{equation}
\label{external flow}
\bv^\infty(\br)=\dot\gamma\bE\cdot \br,
\end{equation} 
where $\dot\gamma$ is the strain rate, and
$\bE$ is the velocity gradient tensor. Linear flows are defined by
\begin {equation}
\label{velocity gradient}
\bE=\half \left( \begin{array} {ccc}
     0& 1+\omega & 0\\
     1-\omega& 0 & 0\\
     0&0&0
\end{array} \right),
\end{equation}
 where $\omega$ is the magnitude of the rotational flow component. Irrotational flow such as the pure extensional flow is given by $\omega=0$. Simple shear flow is specified by $\omega=1$, i.e. $v^{\infty}_x=\dot \gamma y$. A sketch of the problem is shown in Figure~\ref{sketch}.

Typically vesicles are micron-sized. At these small length-scales water is effectively very viscous, inertial effects are unimportant and  fluid velocity fields inside 
$\bv^{\ins}$  and outside $\bv^{\out}$ the vesicle are described by the Stokes equations
\begin{equation}
\label{Stokes equations}
\begin{split}
\eta \nabla^2\bv^\out-\bnabla p^\out
  =0,& \quad \bnabla\cdot\bv^\out=0\,, \\
(\visrat -1)\eta \nabla^2\bv^\ins-\bnabla p^\ins
  =0,& \quad\bnabla\cdot\bv^\ins=0\,.
\end{split}
\end{equation}
Far away from the vesicle, the flow field tends to the unperturbed
external flow $\bv^{\out}\rightarrow \bv^{\infty}$. Velocity is
continuous across the interface
\begin{equation}
 \bv^{\ins}=\bv^{\out}\equiv\bv_s  \quad \mbox{at}\,\quad r=r_s\,,
\end{equation}
where $r_{\mathrm{s}}$ denotes the position of the interface.
The interface moves with the fluid velocity \cite{Leal:1992}
\begin{equation}
\label{interface evolution}
\frac{\partial F}{\partial t}+\bv_{\mathrm{s}}\cdot\bnabla F=0\,,
\end{equation}
where $\bv_{\mathrm{s}}$ is the fluid velocity at the vesicle interface.
Fluid motion gives rise to bulk hydrodynamic stress  
\begin{equation}
\label{stress definition}
\bT=\textstyle -p\bI+\eta(\bnabla\bu+(\bnabla\bu)^T)\,,
\end{equation}
where $\bI$ denotes the unit tensor and the superscript ${\it T}$ denotes transpose. The surrounding fluids exert tractions on the membrane that are balanced by membrane forces
\begin{equation}
\label{stress balance}
\bn\cdot\left(\bT^\out-\bT^\ins\right)=\bt^{\mathrm{mem}} \quad \mbox{at}\,\quad r=r_s\,
\end{equation}
where $\bn$ is the outward unit normal vector and the membrane surface forces are discussed next.

\subsection{Membrane mechanics}

Unlike drops, which are governed by surface tension, the shape of the
vesicle  is determined by bending elasticity\cite{Seifert:1997}.  A
nonequilibrium membrane configuration gives rise to a surface force
density\cite{Seifert:1999}
\begin{equation}
\label{interfacial stress}
\begin{split}
\bt^{\mathrm{mem}}=&\left[2\sigma H  -\kappa \left(4H^3-4KH+2\nabla_s^2 H\right)\right]\bn\\
&-\bnabla_s \sigma\,, 
\end{split}
\end{equation}
where $\kappa$ is the  bending rigidity,  $H$ and $K$ are the mean and Gaussian curvatures, respectively, and
$\sigma$ is the local membrane tension. The tangential part of the surface force density arises from nonuniformities in the surface tension, which are needed to ensure local area incompressibility, as discussed in more details in the next section.
The mean and Gaussian curvatures are 
\begin{equation}
\label{mean curvature}
H=\frac{1}{2}\bnabla\cdot \bn\,,
\end{equation}
\begin{equation}
\label{Gaussian}
K=\frac{1}{2}\bnabla\cdot\left(\bn \bnabla\cdot \bn+\bn\times\left(\bnabla\times\bn\right)\right)\,,
\end{equation}
where the outward unit normal vector can be determined  from the shape function \refeq{F} via 
\begin{equation}
\label{normal vector}
\bn=\frac{\bnabla F}{|\bnabla F|}\,.
\end{equation} The surface gradient operator
is defined as
\begin{equation}
\label{surface gradient}
\bnabla_{\mathrm{s}}=\bI_\mathrm{s}\cdot\bnabla\,,
\end{equation}
where the matrix $\bI_\mathrm{s}=\bI-\bn\bn$ represents a surface projection.
At rest, Eq. \refeq{stress balance} reduces to the well--known Euler--Lagrange equation
\begin{equation}
\label{Euler-Lagrange}
p^\ins-p^\out=2\sigma H -\kappa \left[4H^3-4KH+2\nabla_s^2 H\right]\,.
\end{equation}

\subsection{Area constraint and tension}

The lipid bilayer is fluid because the lipids can diffuse rapidly
within the membrane.  Moreover, since the number of lipids in a
monolayer and the area per lipid are fixed, the lipid bilayer membrane
is incompressible and the total area is conserved.  A membrane element
only deforms but does not change its area.  Accordingly, the local
tension changes in order to keep the local area constant. Hence,
inhomogeneities in the tension ensure local area incompressibility.
The situation is analogous to three--dimensional incompressible
fluids, where finite changes in pressure (tension is the
two-dimensional analogue) correspond to infinitesimal changes in fluid
density, and pressure takes the place of the density as an independent
field variable; in flowing fluids pressure can become nonuniform. The
local area conservation implies that the velocity field at the
interface is solenoidal\cite{Seifert:1999}
\begin{equation}
\label{membrane incompressibility}
\bnabla_s \cdot \bv=0\,.
\end{equation}

The global area constraint acts like an isotropic tension whose value
at equilibrium is given by the Lagrange multiplier used to determine
the shape\cite{Seifert:1995}. Under flow,  the changes in shape result in variations of this  effective isotropic tension.

\subsection{Dimensionless parameters}

Viscous forces exerted by the extensional component of the external
flow drive shape deformation that occurs on a time scale
\begin{equation}
\label{deformation time:3}
t_\deform=\visrat\dot\gamma^{-1}\,.
\end{equation}
Several intrinsic relaxation 
mechanisms oppose the deformation.
  Bending
stresses work to bring the shape back to its preferred curvature
state; the corresponding time scale is
\begin{equation}
\label{Be-relaxation time}
t_{\kappa}=\frac{\visrat\eta a^3}{\kappa}\,.
\end{equation}
In shear flow, vesicle rotation away from the extensional axis of the imposed flow effectively decreases  the extent of shape distortion;  the associated  time scale is
\begin{equation}
\label{rotation}
t_\rot=\dot\gamma^{-1}.
\end{equation}
The strength of these two relaxation mechanisms that
limit  shape deformation by the flow 
is quantified  by the corresponding dimensionless parameters:  
the capillary number 
\begin{equation}
\label{capillary number:bending}
\chi=\frac{t_{\kappa}}{t_\deform},
\end{equation}
and the rotation parameter
\begin{equation}
\label{rotation number}
\frac{t_\rot}{t_\deform}=\visrat^{-1}.
\end{equation}

The smaller of these parameters controls the magnitude of vesicle
deformation.  
At moderate viscosity ratios, the vesicle shape should remain
close to the equilibrium one provided that the capillary number is
small, i.e. restoring bending forces are stronger than the distorting viscous forces.  
A high viscosity inner fluid limits the shape distortion in flows with
non-zero vorticity by means of increasing the rate of vesicle
rotation. In addition to the flow--related parameters, a physical
parameter that arises from vesicle geometry is the excess area
\refeq{excess area}, which sets the maximum magnitude of the shape
distortion; it turns out to be the relevant small parameter for the
analysis of nearly--spherical vesicles as shown in the next sections.

Henceforth, bending stresses and tension are normalized by $\kappa/a^2$; all other quantities are
rescaled using $\eta$, $a$, and $\dot\gamma$.  Accordingly, the time
scale is $\dot\gamma^{-1}$, the velocity scale is $\dot\gamma a$, bulk
stresses are scaled with $\eta\dot\gamma$.

\section{Small deformation theory}
In this section we present a perturbative solution for the
microhydrodynamics of a vesicle with a nearly spherical shape. 

In the coordinate system centered at the vesicle, 
the radial position  $r_{\mathrm{s}}$ of the vesicle
interface can be represented as
\begin{equation}
\label{perturbation of shape}
r_{\mathrm{s}}=\alpha+f\brac{\Omega}\,,
\end{equation}
where $f$ is the deviation of
vesicle shape from a sphere, which depends only on the angles ($\theta\,, \phi$) (or equivalently the solid angle $\Omega$) and
has a vanishing angular average 
\begin{equation}
\label{f:ang average}
\int f\, \dif{\Omega}=0\,.
\end{equation}
The isotropic contribution
$\alpha$ is determined by the volume-conservation constraint
\begin{equation}
\label{volume constraint}
\int (\alpha+f)^3\, \dif{\Omega}=4\pi\,.
\end{equation}
The  total  area conservation  constraint relates the amplitude of the perturbation $f$ and the excess area $\Delta$
\begin{equation}
\label{area constraint}
A/a^2=\int \frac{r_s^2}{\hat\br\cdot\bn}d \Omega=4 \pi+\Delta\,.
\end{equation}
where $\hat\br$ denotes the unit radial vector.

\subsection{Expansion in spherical harmonics}

In Eq.\ \refeq{perturbation of shape}, the function $f$
representing the perturbations of the vesicle shape depends only on angular coordinates.  Thus, it is expanded into series of scalar spherical harmonics
$Y_{jm}$ \refeq{normalized spherical harmonics}
\begin{equation}
\label{expansion of shape in harmonics}
f=\sum_{j=2}^{\infty}\sum_{m=-j}^j f_{jm} Y_{jm}\,,
\end{equation}
In the above equation, the summation starts from nonzero $j$ because
$f$ includes only the nonisotropic contributions.
For small shape perturbations around a sphere the volume constraint
\refeq{volume constraint} becomes\cite{Seifert:1999, Vlahovska:2005}
\begin{equation}
\label{alpha to second order}
\alpha=1- \frac{1}{4\pi}\sum_{jm}f_{jm}f^*_{jm}+O(f^3)\,,
\end{equation}
where the sum over $j$ starts from 2, $|m|\le j$ and $f^*_{jm}=(-1)^mf_{j-m}$.
Similarly, the area constraint  \refeq{area constraint} transforms to
\begin{subequations}
\label{delta to second order}
\begin{equation}
\Delta=\sum_{jm}a(j)f_{jm}f^*_{jm}+O(f^3)\,,
\end{equation}
where
\begin{equation}
a(j)=\frac{\left(j+2\right)\left(j-1\right)}{2}\,.
\end{equation}
\end{subequations}

The  tension is expanded in scalar spherical harmonics
\begin{equation}
\label{expansion of tension in harmonics}
\sigma=\sigma_0+\sum_{j=2}^{\infty}\sum_{m=-j}^j \sigma_{jm} Y_{jm}\,,
\end{equation}
where $\sigma_0$ is the isotropic part of the tension, which varies
with shape in order to keep the total area constant. It is determined
from the condition that the modes must satisfy the area constraint $\dot
\Delta=0$, i.e.
\begin{equation}
\sum_{jm} a_j\dot f_{jm}f^*_{jm}=0\,.
\end{equation}
The nonuniform part of the tension is related to the local incompressibility.

\subsection{Perturbation solution}
The combination of perturbative analysis and the spherical harmonics formalism for solving Stokes-flow problems involving a deformable particle has been developed in detail in Vlahovska {\it{et al.}} \cite{Vlahovska:2003, Vlahovska:2005} for the problem of a surfactant--covered drop. Here we present a brief outline of the method, and focus on the new feature specific to lipid bilayer membranes which is the area constraint.

This paper considers a non--spherical particle albeit the shape
deviation from sphericity is small, i.e.  $f\sim O(\eps)$ and $\eps\ll
1$. In this case, the exact position of the interface
is replaced by the surface of a sphere of equivalent volume, 
and all quantities that are to be evaluated at the interface of the
deformed particle are approximated using a Taylor series expansion.

\subsubsection{Small parameter for area-constrained dynamics}

Small deformations of initially spherical drops and capsules  with elastic membranes have been considered in a number of studies \cite{Barthes_Biesel-Acrivos:1973,Barthes_Biesel:1980}. Typical choices for the small parameter are the capillary number or the inverse viscosity ratio.

In contrast to drops and capsules, the rest shape of the vesicle we
consider is non--spherical. The excess area plays a crucial role in the vesicle
dynamics, because under the constraints of constant area and volume a
sphere is a geometrically rigid object. Consequently, in the
absence of excess area, a vesicle  behaves as a
rigid sphere.

Thus, the appropriate small parameter that reflects the importance of the excess area in the vesicle dynamics is 
\begin{equation}
\eps=\Delta^{\half} \,.
\end{equation}
The square root comes from the observation that $f^2\sim\Delta$~\refeq{delta to second order}.  The choice of the excess area as the small
parameter allows vesicle behavior to described, where the whole excess
area is involved in the vesicle deformation.

\subsubsection{Evolution equation}

We solve the hydrodynamic problem to obtain the velocity field, and
use the fact that the interface moves with the normal component of the
velocity to determine the shape evolution.  The solution for small
deviations from a sphere is presented in details in Appendix
\ref{leading sol}. In this section we present the general expression
for the shape evolution equation valid for any linear flow. In the
subsequent sections we analyze in more details the particular case of
simple shear flow.

At leading order the evolution of the  shape parameters \refeq{expansion of shape in harmonics} is described by
\begin{widetext}
\begin{equation}
\label{ev equation f}
\dot f_{jm}=\im\omega \frac{ m}{2}f_{jm}+\visrat^{-1}C(\visrat, j,m)+\visrat^{-1}\chi^{-1}\Gamma\left(\visrat, \sigma_0, j \right)f_{jm}+O\left(\visrat^{-1}\eps, \eps^2\right)\,,
\end{equation}
\end{widetext}
where the isotropic tension $\sigma_0$ is determined using the area constraint
\refeq{delta to second order} as described in Appendix \ref{Area
constraint}, see Eq. \refeq{tension}. The first term in the evolution
equation \refeq{ev equation f} describes rigid body rotation of a
particle with shape $f$ by the rotational component of the external
flow.  The second term describes the distortion of the vesicle shape
by the extensional component of the external flow. The term including
$\chi$ is associated with relaxation driven by the membrane
stresses. The expressions for $C$  and $\Gamma$ are given by \refeq{coefs in ev eq}. These
coefficients depend on $\visrat$  and are bounded at $\visrat\rightarrow
\infty$.  

The area constraint couples all modes,  and results in isotropic tension, which depends nonlinearly on the shape.
Thus the leading order
vesicle dynamics is non--linear in contrast to the corresponding results for drops and capsules.  As we show in the next section, a
peculiar consequence from the area--constrained dynamics is that the
stationary solution is { {independent}} of the capillary number.



\section{Results: Simple shear flow}

In this section we provide analytical solutions for the vesicle shape
evolution in a simple shear flow.  Simple shear flow consists of an  extensional component, $\bv^\infty_{ext}=\half(y,x,0)$ , which in the
spherical harmonics representation \refeq{velocity fields} is described by
\begin{subequations}
\label{shear flow}
\begin{equation}
\label{extensional part}
\textstyle c^{\infty}_{2\pm20}=\mp \im \sqrt{\frac{\pi}{5}}\,\quad c^\infty_{2\pm22}=\mp \im \sqrt{\frac{2\pi}{15}} \,,
\end{equation}
and  a rotational component, $\bv^\infty_{rot}=\half(y,-x,0)$, which is specified by
\begin{equation}
\label{rotational part}
\textstyle c^\infty_{101}=\im \sqrt{\frac{2\pi}{3}}\,.
\end{equation}
\end{subequations}
The extensional part of external flow \refeq{external flow}, which is responsible for shape distortion, is fully
characterized by a second-order traceless tensor, which corresponds to
spherical harmonics of the order $j=2$ \refeq{extensional part}. 
Therefore, at leading order the shear flow affects only  the subspace $j=2$. 
Considering only these modes simplifies the expression for the tension  \refeq{tension} and  the evolution equations \refeq{ev equation f}  to the following  set of coupled non--linear differential equations
\begin{equation}
\label{f22 evolution}
\begin{split}
\dot f_{2m}=&\textstyle -\frac{\im m}{2} h(\visrat)\delta_{|m|2}+\frac{\im m}{2}f_{2m}\\
&-2\im h(\visrat) \Delta^{-1}(f_{22}-f_{2-2})f_{2-m}
\end{split}
\end{equation}
where 
\begin{equation}
h(\visrat)=\frac{4 \sqrt{30 \pi}}{23\visrat+9}\,.
\end{equation} 
Strictly speaking Eq. \refeq{f22 evolution} is only valid  at long times ($t\gg t_\kappa $) when all transient $j\neq 2$ modes have decayed.
The initial conditions are set by the equilibrium vesicle configuration. For example, for a fluctuating quasi--spherical vesicle  \cite{Seifert:1999} 
\begin{equation}
|f_{jm}|^2=\frac{k_BT}{\kappa  E(j, \sigma_\eq)}\,,
\end{equation}
where $E(j,\sigma)$ is given by \refeq{Ej}.

The shape  parameters \refeq{expansion of shape in
harmonics} are decomposed
into real and imaginary parts
\begin{equation}
f_{jm}=f'_{jm}+\im f''_{jm}\,\,\quad f^*_{jm}=f'_{jm}-\im f''_{jm}
\end{equation}
In the flow plane $x-y$, the vesicle shape $f$ is characterized by only three components, $f'_{22}$ ,$f''_{22}$ , and $f_{20}$, corresponding to deformation along the flow  axis $x$, straining axis $x=y$, and the $z$ axis. Shape modes $f_{2\pm1}$ describe deformations of the type $xz$ and $yz$, which vanish in the flow plane $z=0$. From geometrical considerations we have  for the  inclination angle 
\begin{equation}
\label{angle: shear} 
\phi_0=-\frac{1}{2}\arctan\left(\frac{f_{22}''}{f_{22}'}\right)\,.
\end{equation}
\subsection{Tank--treading}
\begin{figure}
\epsfxsize=3in
\epsffile{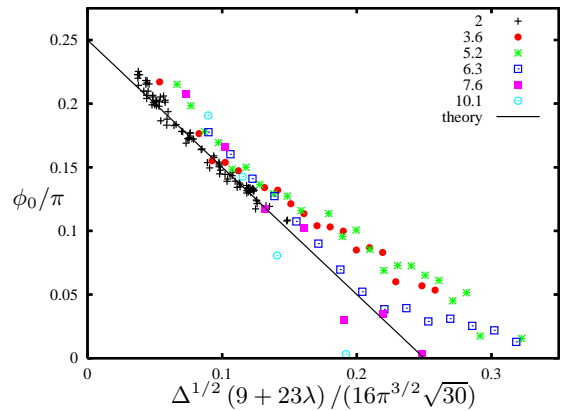}
\begin{picture}(0,0)(0,0)
\put(-220,70){$\phi_0/\pi$}
\put(-160,-5){$\Delta^{1/2}\left(9+23\visrat\right)/(16\pi^{3/2}\sqrt{30})$}
\end{picture}
\caption{\footnotesize Stationary inclination angle in simple shear flow as 
function of the rescaled excess area  for different viscosity contrasts. The symbols are experimental data from Kantsler and Steinberg \cite{Kantsler-Steinberg:2006}. The lines represent the linear small
deformation theory \refeq{angle for small Delta}.}
\label{kantsler}
\end{figure}

The evolution equations \refeq{f22 evolution} have two sets of stationary points. The first one corresponds to the tank--treading state. 
The only non--zero stationary amplitudes are
\begin{equation}
\label{amplitudes:1}
f_{22}''=-\frac{\Delta}{4h(\visrat)}\sqrt{-1+\frac{4h^2(\visrat)}{\Delta}}\,,\quad f_{22}'=\frac{\Delta}{4 h(\visrat)}\,.
\end{equation}
The numerical solution of the full set of  evolution equations \refeq{ev equation f} shows that all other modes decay to zero. Hence, in the stationary state   all excess area is stored in the  $f_{2\pm2}$ modes.
Substituting the stationary amplitudes \refeq{amplitudes:1} in the relation for the inclination angle \refeq{angle: shear} and expanding  for small values of the excess area $\Delta$
we obtain
\begin{equation}
\label{angle for small Delta}
\phi_0=\frac{\pi}{4}-\frac{(9+23 \visrat)\Delta^{1/2}}{16\sqrt{30 \pi}}\,,
\end{equation}
which agrees with the expression reported by Seifert \cite{Seifert:1999} for the case of no viscosity contrast ($\visrat=2)$. The analogous result reported by Misbah \cite{Misbah:2006} contains a misprint(140 should read 240).
Our relation \refeq{angle for small Delta} is in good quantitative agreement with the experimental data of Kantsler and Steinberg \cite{Kantsler-Steinberg:2005}, as demonstrated in Figure \ref{kantsler}.
Eq.\ \refeq{amplitudes:1} implies that if $\Delta> 4h^2$, i.e. if the viscosity contrast is higher than a critical value
\begin{equation}
\label{crit viscosity}
\visrat_c=-\frac{9}{23}+\frac{120}{23}\sqrt{\frac{2\pi}{15 \Delta}}
\end{equation}
no stationary tank--treading solution exists, as also pointed out by Misbah\cite{Misbah:2006, Biben-Misbah:2003}, and the vesicle starts to tumble. 

\subsection{Tumbling}

The second fixed point of Eq. \refeq{f22 evolution} corresponds to periodic vesicle deformation given by oscillating $j=2$ modes around  values
\begin{equation}
f_{22}'=h(\visrat)\,,\quad f_{22}''=0\,\quad f'_{21}=0\,,\quad f''_{21}=0\,. 
\end{equation}
If $f_{20}(0)\neq 0$, the $f_{20}(t)$ is also oscillating.
All modes with $j\ne 2$ decay to zero
and are not oscillating.

\begin{figure}
\epsfxsize=3in
\epsffile{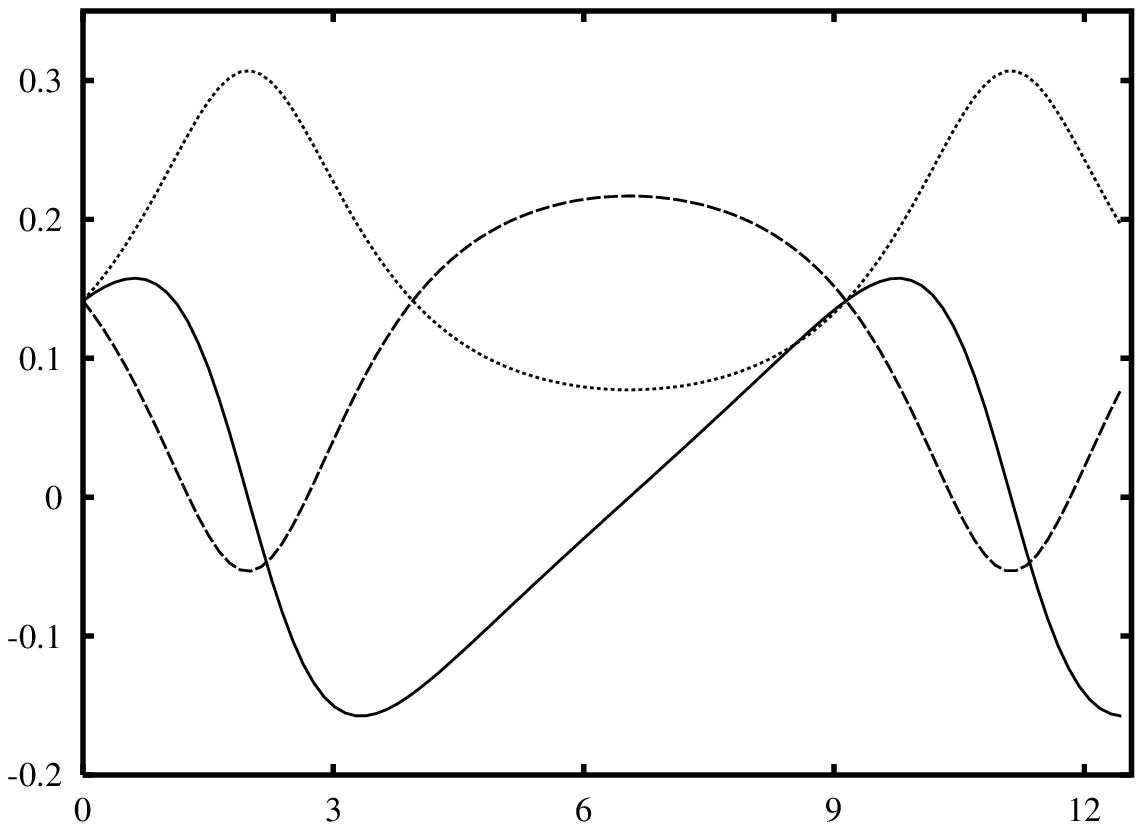}
\epsfxsize=3in
\epsffile{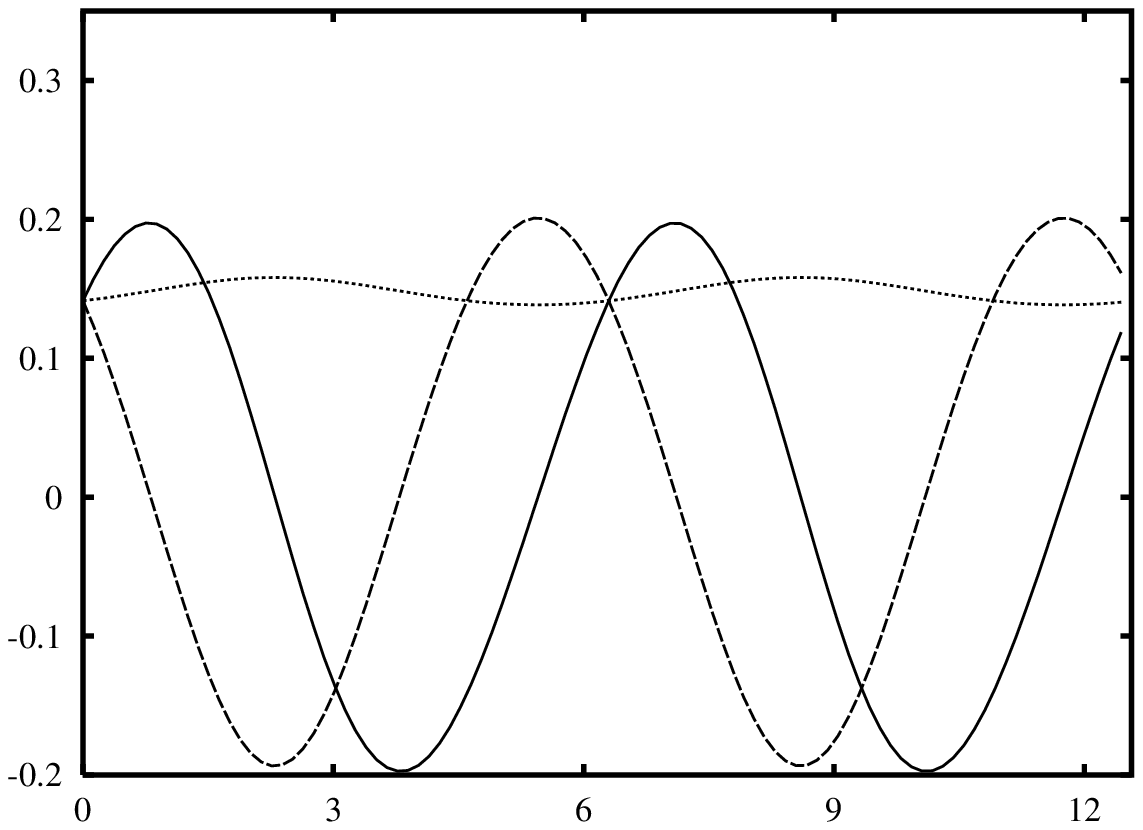}
\begin{picture}(0,0)(0,0)
\put(-290,130){(a)}
\put(-190,130){(b)}
\put(-220,220){$f_{2m}$}
\put(-220,70){$f_{2m}$}
\put(-120,-5){$t$}
\end{picture}
\caption{\footnotesize Time dependence of the $f_{2m}$ modes in the (a) breathing ($\visrat\gtrsim \visrat_c$) and (b) tumbling ($\visrat\gg \visrat_c$) regimes; solid, dashed and dotted lines represent the $f_{22}'',\, f_{22}'$ and $f_{20}$ modes, respectively. The excess area is  $\Delta=0.2$. The viscosity contrasts are  (a) $\visrat=10$,(b) $\visrat=100$. The initial conditions are $f'_{22}(0)=f''_{22}(0)=\sqrt{0.1\Delta}$; $f_{20}(0)$ is determined by  the area constraint. }
\label{osc}
\end{figure}

The time--periodic vesicle deformation depends strongly on the
viscosity contrast as illustrated in Figure \ref{osc}.  Figure
\ref{osc} (a) shows that close to the critical viscosity contrast the
period of oscillations is long and the amplitude of oscillations of
the $f_{20}$ mode, which corresponds to out-of-the-flow-plane
deformation, is significant.  Experimentally the vesicle
appears to be trembling or ``breathing''
\cite{Kantsler-Steinberg:2006,Mader:2006}. The ``breathing'' motion
has been discussed in parts by Misbah \cite{Misbah:2006} although the
role of the $f_{20}$ mode has not been recognized. 

The mode
oscillation frequency increases with viscosity contrast, as seen in
Figure \ref{osc} (b).  At high viscosity contrast it approaches twice the
 rate of rotation of the external flow, $\dot\gamma$.  This can
be seen from Eq. \refeq{f22 evolution} where in the limit $\visrat\gg
1$ only the rotation terms survive.  Since the $f_{20}$ mode is not
rotationally stabilized, its oscillation amplitude becomes negligible
at high viscosity contrast; correspondingly the out-of-the-flow-plane
deformation is suppressed. The periodic vesicle deformation then
corresponds to rigid body rotation, $f'_{22}(t)\sim \cos(t)\,, \quad
f''_{22}(t)\sim\sin(t)$.

In order to compare theory and experiment we  derive expressions for some experimentally measurable parameters. 
Experimentally only the deformation in the flow plane is
observable. The vesicle contour represents an ellipse.   Combining \refeq{perturbation of shape} and \refeq{angle: shear} we obtain for the lengths of the major and minor axes 
\begin{equation}
r_{max, min}(t)=\textstyle{1-f_{20}(t)\sqrt{\frac{5}{\pi}}\pm\left(\Delta-2f_{20}^2(t)\right)^\half\sqrt{\frac{15}{2\pi}}}\,,
\end{equation}
where we have neglected the contributions from the $f_{2\pm1}$ modes
for the sake of simplicity.  We see that the major and minor axes
follow the oscillations of the $f_{20}$ mode: the larger the $f_{20}$
oscillation amplitude, the larger the amplitude in the $r_{max}$ and
$r_{min}$ fluctuations.  In the experiments of Kantsler and
Steinberg\cite{Kantsler-Steinberg:2006}, the excess area $\bar\Delta$
corresponding to a prolate ellipsoidal shape with major and minor axes
as seen in the flow plane was reported.  Using the relation between
the deformation in the flow plane and the excess area
$(r_{max}-r_{min})/(r_{max}+r_{min})=(15 \bar\Delta/32\pi)^\half$,
reported first by Seifert\cite{Seifert:1999}, we obtain
\begin{equation}
\label{Delta_{xy}}
\bar \Delta(t)=\frac{\Delta-2 f^2_{20}(t)}{\left[1-f_{20}(t)\sqrt{\frac{5}{16\pi}}\right]^2}\,.
\end{equation}
Unlike the true excess area $\Delta$,
which is constant, $\bar\Delta$ is time--dependent due to the
fluctuations in the $f_{20}$ mode, i.e. the out-of-the-flow-plane
transfer of area.

The angle evolution is described by
\begin{equation}
\label{angle evolution}
\dot \phi_0=-\frac{1}{2}+\frac{h(\visrat)}{(\Delta-2 f_{20}^2(t))^\half}\cos(2 \phi_0(t))\,, 
\end{equation}
from which we infer that the tumbling period, defined as the time needed for a material point to return to its initial position, is given by 
\begin{equation}
\label{tumbling period}
T_{tumble}=4 \pi\left[1-\frac{4h(\visrat)^2}{\Delta}\right]^{-\half}\,.
\end{equation}
As already discussed, when the viscosity contrast is high the
oscillations of the $f_{20}$ mode are suppressed and $f_{20}(t)\approx f_{20}(0)$. In this case Eq.\ \refeq{angle evolution} becomes equivalent to the Keller--Skalak equation $\dot \phi_0=A+B\cos(2 \phi_0)$, where  the constants $A$ and $B$ are identified  with $-1/2$ and $h(\visrat)/\sqrt{\Delta-2f_{20}^2(0)}$, respectively.

We compared the time--dependent vesicle behavior with the experiments
of Mader {\it{et al.}}\cite{Mader:2006}. A good agreement was obtained
for the angle evolution, for example of their vesicle number 6 using
reduced volume 0.988 instead of the reported 0.996. The difference is
reasonable given the uncertainty with which the reduced volume is
determined experimentally: it is computed assuming axial symmetry
about the vesicle's longest axis, which might not be always the case
because of the presence of a non-zero $f_{20}$ mode. Note, however, that in the
tank--treading regime, since the excess area is stored only
in the $f_{2\pm2}$ modes the vesicle is a prolate ellipsoid and the
computation of the reduced volume is quite accurate.  We also
attempted  to reproduce theoretically the experimental curves in Figure 4 of
Kantsler and Steinberg\cite{Kantsler-Steinberg:2006}. The predicted
tumbling frequency from Eq. \refeq{tumbling period} using viscosity
contrast and excess area close to the reported values was almost twice
those of the experimental data. One possible explanation is that
our small--deformation theory fails for the large values of excess
area in this experiment ($\Delta\approx 1$).

\subsection{Rheology}

A suspension of vesicles can be described as  a continuum with
effective properties at length scales large compared to the size of the constituent particles.  In dilute suspensions
particles are far from each other and they do not feel each
others presence. Thus, the effective stress is just a sum of the
stress contributions of the individual particles and the bulk stress
of a dilute suspension is linear with the particle volume fraction
$\varphi$ \cite{Kim-Karrila:1991, Pozrikidis:BIM}
\begin{equation}
\bm \Sigma=2 \bE^s+\varphi \bT\,,
\end{equation}
where $ \bE^s$ is the symmetric part of the velocity gradient tensor \refeq{velocity gradient} and $\bT$ is the stress associated with the vesicle disturbance
velocity field.  Rheological properties of interest are the particle
contribution to the shear viscosity, $T_{xy}$, and the normal stress
differences, $N_1=T_{xx}-T_{yy}$ and $N_2=T_{yy}-T_{zz}$.
Vesicle stress is
directly related to the amplitudes of the velocity field $j=2$
\refeq{velocity fields} \cite{Blawzdziewicz-Vlahovska-Loewenberg:2000}  corresponding to the stresslet (the symmetric force dipole).
Using the relations \refeq{stresses:Multipoles} we obtain for 
the vesicle contribution to
the effective shear viscosity 
\begin{equation}
\label{shear viscosity:1}
T_{xy}=\textstyle{\frac{5}{2}\frac{23\visrat-39}{23\visrat+9}+\chi^{-1}3\im\sqrt{\frac{30}{\pi}}\frac{E(2, \sigma_0)\left(f_{22}-f_{2-2}\right)}{23\visrat+9}\,.}
\end{equation}
where the shape deformation modes $f_{2\pm2}$ are given by
\refeq{amplitudes:1app}. 
For a given excess area and viscosity
contrast
the effective viscosity in the tank--treading regime 
is 
\begin{equation}
\label{shear viscosity}
T_{xy}=\frac{5}{2}-\Delta\frac{(23\visrat+9)}{16\pi}\,,
\end{equation}
which is obtained by inserting the expression for the modes amplitude
\refeq{amplitudes:1app} in Eq. \refeq{shear viscosity:1} and taking into
account \refeq{prolate}. In the limit of a spherical particle,
$\Delta=0$, and since a sphere with fixed area and volume in shear flow undergoes only a rigid body rotation, the Einstein's result for a suspension of rigid spheres is
recovered.  The effective viscosity decreases
with the increase of the excess area because the deformable vesicles
elongate and thus offer less resistance to the flow. An increasing
viscosity contrast also leads to a decrease in the effective viscosity
because vesicles align better with the flow (the inclination angle
\refeq{angle for small Delta} decreases with viscosity contrast).

Unlike a suspension of rigid spheres, a suspension of deformable vesicles exhibits normal stresses. In the tank--treading regime below the critical viscosity contrast,  the magnitudes of the first normal stress difference is
\begin{equation}
\label{norm stress}
N_1= \frac{\Delta(23 \visrat+9)}{8 \pi}\left[-1+\frac{1920\pi}{\Delta(23 \visrat+9)^2}\right]^\half\,,
\end{equation}
and the second normal difference is $N_2=-\half N_1$.

\subsection{Vesicle migration in the presence of a wall}

A spherical particle in simple shear flow produces a symmetric
disturbance velocity field and therefore the particle does not drift
relative to a bounding wall \cite{Leal:1980}. Particle deformation
breaks the symmetry and may lead to cross--stream migration. 

The leading order term in the far field of the disturbance velocity
due to a force--free and torque--free particle is the stresslet, the
symmetric and traceless force dipole. The boundary conditions at the
wall can be satisfied by placing the stresslet hydrodynamic image on
the opposite side\cite{Kim-Karrila:1991}. Thus, a particle far away
from the wall moves with a velocity due to its corresponding image
stresslet; in particular, the vesicle drift velocity normal to a rigid
wall is proportional to the stresslet component in the direction of
the plane unit normal, i.e. $T_{yy}$. Smart and Leighton
\cite{Smart-Leighton:1990} report that
\begin{equation}
U_{\mathrm{lift}}=-\frac{3}{16d^2} T_{yy}\,,
\end{equation}
where $d$ denotes the distance from the particle center to the wall.
Taking into account that the normal stress is $T_{yy}=-(N_1-N_2)/3$
and using the normal stresses expressions \refeq{norm stress} we
obtain for the lift velocity
\begin{equation}
U_{\mathrm{lift}}=\frac{c(\visrat, \Delta) }{d^2}\,,
\end{equation}
where 
\begin{equation}
\label{prefactor}
c(\visrat, \Delta)=\textstyle \frac{3}{32} N_1\,.
\end{equation}
Similar type expressions have been reported in  analytical studies \cite{Olla:1997a, Seifert:1999b},
although the dependence on the excess area has not been presented explicitly. Eq.~\refeq{prefactor} shows that the more deformable the vesicle, i.e. the larger the
excess area, the larger the migration velocity. Interestingly, at very small excess area we obtain from \refeq{norm stress} that $N_1\sim~\Delta^\half$, and vesicle migration is independent of the viscosity contrast.  The values of the
prefactor \refeq{prefactor} are of the same order of magnitude as
reported in studies considering a tank-treading ellipsoid
\cite{Olla:1997a} or numerical simulations of a vesicle with no
viscosity contrast \cite{Sukumaran-Seifert:2001}.

\section{Summary and outlook}

This study considers the behavior of a vesicle formed by a fluid
membrane in a general shear flow.  The analysis takes into account the
membrane incompressibility and bending resistance as well as the 
 viscosity contrast between the interior and exterior fluids.
Analytical results for a nearly spherical vesicle are obtained.
Expressions for the shape evolution equation, stress and velocity
fields and the effective membrane tension, which is needed to enforce
the area constraint, were derived using a formalism based on spherical
harmonics.  The main theoretical result of this study is contained in
the shape evolution equation Eq.\ \refeq{ev equation f} and the
expression for the effective tension Eq.\ \refeq{tension}. The latter
serves to eliminate the tension in favor of the excess area, which is
the physically relevant parameter.  In contrast to drops and capsules,
the shape evolution of the area--constrained vesicle is non--linear.

The derived shape evolution equation describes vesicle dynamics in
general linear flow.
In the particular case of simple shear flow, only the shape modes with the same
spherical harmonic order as the external flow, $j=2$, are perturbed;
all other modes decay on a time scale set by the bending
rigidity. Consequently, the shape evolution equations simplify to
Eq.~\refeq{f22 evolution} and yield expressions for quantities that
can be measured experimentally such as the inclination angle with
respect to the flow direction, the lengths of the major and minor axes
of the vesicle contour in the flow plane, etc.  In the stationary
tank--treading state, we show that in a simple shear flow the leading
order deformation and stresses are independent of the membrane bending
rigidity.  Our theory is in quantitative agreement with experimental
data for the vesicle deformation in the tank--treading and in the
tumbling regimes.  Non--Newtonian rheology with normal stresses is
predicted for a suspension of vesicles. We also considered the vesicle
dynamics in wall--bounded shear flow and presented a simple derivation
of the leading order correction, in the distance to the wall, to the
rate of vesicle migration away from the wall.

Several  problems remain open. 
We discussed the main features of the time--dependent vesicle dynamics
in linear flows, but this problem remains to be systematically
explored.  Although the quantitative agreement between predicted and
experimentally observed\cite{Mader:2006}  behavior of tumbling vesicles with small
excess area is encouragingly good, for vesicles with large excess area 
\cite{Kantsler-Steinberg:2006} the agreement is poor,
 which might be due to the limitations of the leading order theory.
Our analysis will be extended to consider higher--order
perturbations, and hence elucidate the feedback from the flow on vesicle
dynamics.  

The developed theory applies only to a nearly spherical vesicle. Larger
deformations from sphericity can be explored by numerical simulations.
A body of work exists on capsule dynamics, mostly considering elastic
membranes with no bending resistance \cite{Pozrikidis:1995,
Eggleton-Popel:1998, Lac-Barthes_Biesel:2004}. To our knowledge there
are only a couple of numerical studies that include bending resistance
\cite{Kwak-Pozrikidis:2001, Pozrikidis:2001}.
Area incompressible fluid membranes with bending resistance such as
the lipid bilayer membranes, have been studied only to a limited
extent \cite{Kraus-Seifert_etal:1996, Beaucourt_etal:2004,
Noguchi-Gompper:2005} and some results are conflicting. For example,
Noguchi and Gompper\cite{Noguchi-Gompper:2005} report stationary
tank-treading discocyte shape, while Kraus {\it {et
al.}}\cite{Kraus-Seifert_etal:1996} find only prolate ones. More
efficient and accurate simulations are needed in order to perform a
systematic study on flow induced shape transitions. In order to explore numerically the
dynamics of highly non--spherical vesicle shapes we are
 developing Boundary Integral Method simulations, which combine
an algorithm for adaptive restructuring of the computational grid that
allows resolution of high curvature
regions\cite{Cristini-Blawzdziewicz-Loewenberg:2001d} with the area
constraint \refeq{membrane incompressibility} and interfacial force
density \refeq{interfacial stress}.

\section{Acknowledgments}

RSG acknowledges financial support from the Human Frontier Science
Project. PV thanks Rumy Dimova's group for their the hospitality.
This work has benefited from many stimulating discussions with Thomas
Powers and Jerzy Blawzdziewicz.  The authors thank Reinhard Lipowsky
and Markus Deserno for their comments on the manuscript, Manouk
Abkarian, Vasiliy Kantsler and Victor Steinberg for providing their
experimental data, and Miles Page for proofreading the manuscript.

\appendix

\section{Spherical harmonics}
\label{Harmonics}
For the sake of completeness,  we list the definitions of scalar and vector spherical  harmonics \cite{Jones:1985, Varshalovich:1988}.
The normalized spherical scalar harmonics are defined as 
\begin{equation}
\label{normalized spherical harmonics}
   Y_{jm}\brac{\Omega} = \textstyle \left[\frac{2j+1}{4\pi}\frac{(j-m)!}{(j+m)!}\right] (-1)^m P_j^m(\cos\theta)e^{{\rm i}m\varphi},
\end{equation}
where  $\rhat=\br/r$,  $(r, \theta,\varphi)$
are the spherical coordinates, and $P_j^m(\cos\theta)$ are the Legendre polynomials. 
The vector spherical harmonics are defined as
\begin{subequations}
\label{vector harmonics}
\begin{equation}
\label{vector harmonics:S}
\bS_{jm0}=\left[j\left(j+1\right)\right]^{-\half}r\nabla_{\Omega} Y_{jm}\,,
\end{equation}
\begin{equation}
\bS_{jm1}=-\im \rhat\times \bS_{jm0}\,,
\end{equation}
\begin{equation}
\bS_{jm2}=\rhat Y_{jm}\,,
\end{equation}
\end{subequations}
where $\nabla_{\Omega}$ denotes the angular part of the gradient
operator.  $\bS_{jm0}$ and $\bS_{jm1}$ are tangential, while
$\bS_{jm2}$ is normal to a sphere.
\section{Fundamental set of velocity fields}
\label{Ap:velocity basis}
Following the definitions given in Blawzdziewicz {\it et al.}\cite{Blawzdziewicz-Vlahovska-Loewenberg:2000}, we list the expressions for the functions ${\bu}^\pm_{jmq}\left(r,\Omega\right)$  
\begin{subequations}
\label{vel basis -}
\begin{equation}
\label{-vel 0}
\begin{array}{ll}
\bu^-_{jm0}&={\textstyle\frac{1}{2}}r^{-j}\left(2-j+j r^{-2}\right)\bS_{jm0}\\
&+{\textstyle\frac{1}{2}}r^{-j}\left[j\left(j+1\right)\right]^{1/2}\left(1-r^{-2}\right) \bS_{jm2}\,,
\end{array}
\end{equation}
\begin{equation}
\label{-vel 1}
\bu^-_{jm1}=r^{-j-1}\bS_{jm1}\,,
\end{equation}
\begin{equation}
\label{-vel 2}
\begin{array}{ll}
\bu^-_{jm2}&={\textstyle\frac{1}{2}}r^{-j}\left(2-j\right) \left(\frac{j}{1+j}\right)^{1/2}\left(1-r^{-2}\right)\bS_{jm0}\\
&+{\textstyle\frac{1}{2}}r^{-j}\left(j+(2-j)r^{-2}\right)\bS_{jm2}\,.
\end{array}
\end{equation}
\end{subequations}
\begin{subequations}
\label{vel basis +}
\begin{equation}
\label{+vel 0}
\begin{array}{ll}
\bu^+_{jm0}&= {\textstyle\frac{1}{2}}r^{j-1}\left(-(j+1)+(j +3)r^2\right)\bS_{jm0}\\
&-{\textstyle\frac{1}{2}}r^{j-1}\left[j\left(j+1\right)\right]^{1/2}\left(1-r^2\right)\bS_{jm2}\,,
\end{array}
\end{equation}
\begin{equation}
\label{+vel 1}
\bu^+_{jm1}=r^{j}\bS_{jm1}\,,
\end{equation}
\begin{equation}
\label{+vel 2}
\begin{array}{ll}
\bu^+_{jm2}&={\textstyle\frac{1}{2}}r^{j-1}\left(3+j\right)\left(\frac{j+1}{j}\right)^{1/2}\left(1-r^2\right)\bS_{jm0}\\
&+{\textstyle\frac{1}{2}}r^{j-1}\left(j +3-(j+1)r^2  \right)\bS_{jm2}\,.
\end{array}
\end{equation}
\end{subequations}
On a sphere $r=1$ these velocity fields reduce to
\begin{equation}
\bu^{\pm}_{jmq}=\bS_{jmq}\,.
\end{equation}
Hence, $\bu^{\pm}_{jm0}$ and $\bu^{\pm}_{jm1}$ are tangential,  and $\bu^{\pm}_{jm2}$ is normal to a sphere.
In general, a vector velocity field which is tangential to a surface with normal $\bn$ has an irrotational component  \cite{Blawzdziewicz-Wajnryb-Loewenberg:1999}
\begin{equation}
\bn\cdot\left(\bnabla_s\times\bv^{irr}\right)=0\,,
\end{equation}
and solenoidal component
\begin{equation}
\bnabla_s\cdot\bv^{sol}=0\,.
\end{equation}
On a sphere, the irrotational component is identified with the $q=0$ vector spherical harmonic, and the solenoidal corresponds to the  $q=1$ vector spherical harmonic.

\section{Effective stress of a dilute dispersion}
The disturbance velocity field due a particle can be represented as a
superposition of velocity fields generated by a collection of force
multipoles, by analogy to electrostatics
\cite{Cichocki-Felderhof-Schmitz:1988, Kim-Karrila:1991}.  

The strength of the stresslet, which is the symmetric and traceless force dipole moment, gives the particle contribution to the effective stress of a dilute  dispersion
\begin{subequations}
\label{stresses:Multipoles}
\begin{equation}
\label{shear stress in terms of multipoles}
T_{xy}=- \frac{\im}{8}\sqrt{\frac{6}{5 \pi}}\left(\hat c_{220}- \hat c_{2-20}\right)\,,
\end{equation}
\begin{equation}
\label{normal stress 1 in terms of multipoles}
N_1=-\frac{1}{4}\sqrt{\frac{6}{5 \pi}}\left(\hat c_{220}+\hat c_{2-20}\right)\,,
\end{equation}
\begin{equation}
\label{normal stress 2 in terms of multipoles}
N_2=-\half N_1 +\frac{3}{4}\sqrt{\frac{1}{5 \pi}}\hat c_{200}\,.
\end{equation}
\end{subequations}
The stresslet field is   related to the amplitude of the $j=2$ velocity field;  the relations between $\hat c$ and $c$ are given by 
\begin{equation}
\label{CFS basis}
\begin{array}{ll}
\hat c_{jm0}= &\frac{(2j+1)(2j-1)\left[j(j+1)\right]^{1/2}}{j+1}c_{jm0}\\
&+\frac{j(2j+1)(2j-1)}{j+1}c_{jm2}\,.
\end{array}
\end{equation}
The complete expressions can be found in Blawzdziewicz et al. \cite{Blawzdziewicz-Vlahovska-Loewenberg:2000}.

\section{Velocity fields and hydrodynamics stresses}

Velocity fields are described using basis sets of fundamental solutions of the Stokes equations \cite{Cichocki-Felderhof-Schmitz:1988}, $\bu^\pm_{jmq}$, defined in Appendix~\ref{Ap:velocity basis}:
\begin{subequations}
\label{velocity fields}
\begin{equation}
\begin{split}
\bv^{\out}\brac{\br}=&\sum_{jmq} c^{\infty}_{jmq}\left[\bu^{+}_{jmq}\brac{\br}-\bu^{-}_{jmq}\brac{\br}\right]\\
&+\sum_{jmq}c_{jmq}\bu^{-}_{jmq}\brac{\br}\,,
\end{split}
\end{equation}
\begin{equation}
\bv^{\ins}\brac{\br}=\sum_{jmq}c_{jmq}\bu^{+}_{jmq}\brac{\br}\,.
\end{equation}
\end{subequations}
The hydrodynamic tractions exerted on a surface with a  normal vector $\bn$ are $\bn\cdot\bT$
\begin{equation}
\label{traction definition}
\trac\equiv\bn\cdot\bT=\tau_{jmq}\bS_{jmq}
\end{equation}
 In the particular case of a sphere characterized with a normal vector $\hat\br$, the
viscous tractions are linearly related to the velocity field
\begin{subequations}
\begin{equation}
\tau^\out_{jmq}=\sum_{q'}^{2}c^\infty_{jmq'}\left(\Theta_{q'q}^{+}-\Theta_{q'q}^{-}\right)+\sum_{q'}^{2}c_{jmq'}\Theta_{q'q}^{-}\,
\end{equation}
\begin{equation}
\tau^\ins_{jmq}=\sum_{q'}^{2}c_{jmq'}\Theta_{q'q}^{+}\,
\end{equation}
\end{subequations}
where $\Theta_{q'q}^{\pm}$ are obtained from the  velocity fields \refeq{vel basis -}--\refeq{vel basis +} \cite{Blawzdziewicz-Vlahovska-Loewenberg:2000},
\begin{equation}
\label{tracT0+}
\tracT^{+}_{qq'}\brac{j}=
\left(
\begin{array}{ccc}
2j+1 & 0 & -3\brac{\frac{j+1}{j}}^\half \\
0 & j-1 & 0 \\
-3\brac{\frac{j+1}{j}}^\half & 0 & 2j+1+\frac{3}{j} \\
\end{array}
\right)
\end{equation}
\begin{equation}
\label{tracT0-}
\tracT^{-}_{qq'}\brac{j}=
\left(
\begin{array}{ccc}
-2j-1 & 0 & 3\brac{\frac{j}{j+1}}^\half \\
0 & -j-2 & 0 \\
3\brac{\frac{j}{j+1}}^\half & 0 & -2j-1-\frac{3}{j+1} \\
\end{array}
\right)
\end{equation}
In the derivation of the above expressions we have used that
\begin{equation}
\label{trick:2}
\rhat\cdot(\bnabla\bv+(\bnabla\bv)^T)=r\frac{d}{dr}\brac{\frac{\bv}{r}}+\frac{1}{r}\bnabla\brac{\bv\cdot\br}\,.
\end{equation}

\section{Leading order perturbation solution}
\label{leading sol}

For small deviations from sphericity \cite{Vlahovska:2005}, the $O(\eps)$ expansions for the normal vector and the mean curvature are  
\begin{equation}
\label{normal vector expansion}
\bn\brac{\Omega}=\rhat-f_{jm}\left[j\left(j+1\right)\right]^\half\bS_{jm0}\,\,, 
\end{equation}
\begin{equation}
\label{curvature expansion}
H\brac{\Omega}=\textstyle {1+\frac{1}{2}\left(j+2\right)\left(j-1\right)f_{jm}Y_{jm}\,.}
\end{equation} 
At order $O(\eps)$, in the bending  stresses \refeq{interfacial stress} the terms $H^3$ and $HK$ cancel and  only $\nabla_s^2 H$ remains 
\begin{equation}
\nabla_s^2 H=-\frac{1}{2}j\left(j+1\right)\left(j-1\right)\left(j+2\right)f_{jm}Y_{jm}\,.
\end{equation}
Combining Eqs. \refeq{normal vector expansion} and \refeq{curvature expansion} in the Laplace term of the  membrane stresses \refeq{interfacial stress} yields
\begin{equation}
\sigma H=\textstyle \left(\half\sigma_0\left(j-1\right)\left(j+2\right)f_{jm}+\sigma_{jm}\right)Y_{jm}\,.
\end{equation}
In the above equation the isotropic part has been omitted because it
has no importance to flow dynamics; $\sigma_{jm} f_{jm}$ is neglected
as well as it is a higher order term.  On
a sphere $\nabla_s \sigma$ becomes simply $\sqrt{j(j+1)}
\sigma_{jm}\bS_{jm0}$ according to Eq.~\refeq{vector harmonics:S}.

All modes that contribute to the excess area 
are affected by the flow.  Therefore, we will present results for any
$j$. In this way, the theory can be applied to fluctuating vesicles
and other types of flow, e.g. quadratic (Poiseuille) flow.

The incompressibility condition \refeq{membrane incompressibility} implies that the amplitudes of the velocity disturbance field \refeq{velocity fields} are related
\begin{equation}
c_{jm2}=\frac{1}{2}\sqrt{j(j+1)}c_{jm0}\,.
\end{equation}
At leading order  the stress balance reads
\begin{equation}
\hat\br\cdot \bT^\out-(\visrat-1)\hat\br\cdot\bT^\ins=\bt^\mem \,.
\end{equation}
It gives a relation for the tractions \refeq{traction definition}, which is linear in $j$ and $m$
\begin{equation}
\trac^\out_{jmq} -(\visrat-1)\trac_{jmq}^\ins=\bt^{\mem}_{jmq}\,.
\end{equation}
The tangential  interfacial stress has an ``irrotational'' component 
\begin{equation}
\label{bt tang}
\bt^\mem_{jm0}=-\chi^{-1} \sqrt{j(j+1)}\sigma_{jm}\,.
\end{equation}
The normal component of the membrane stress is 
\begin{equation}
\label{bt normal}
\bt^\mem_{jm2}=\chi^{-1} \left(2\sigma_{jm}+E (j, \sigma_0) f_{jm}\right)\,.
\end{equation}
The tangential balance  \refeq{bt tang}
determines the tension distribution in terms of the shape parameters
\begin{equation}
\begin{array}{ll}
\sigma_{jm}&=-\chi c^{\infty}_{jm0}\frac{2\left(1+2j\right) }{\sqrt{j(j+1)}}+\chi c^\infty_{jm2}\frac{3(2j+1)}{j(j+1)} \\
&+ \chi c_{jm0}\frac{\left(3+\visrat\left(j-1\right)\right)}{2\sqrt{j(j+1)}}\,.
\end{array}
\end{equation}
Finally, the normal stress balance \refeq{bt normal} yields $c_{jm0}$. 

Expanding around sphere, for the linear flow \refeq{external flow}
Eq.~\refeq{interface evolution} yields
\begin{equation}
\label{interface evolution:2}
\frac{\partial f_{jm}}{\partial t}= c_{jm2}+\omega\frac{\im m}{2}f_{jm}\quad {\mbox{at}}\,\, r=1\,.
\end{equation}
The first term represents the motion of the interface due to the normal component of the velocity and the second term describes the rotation of the deformed shape.

Substituting $c_{jm2}$ in Eq.~\refeq{interface evolution:2} yields the evolution equation \refeq{ev equation f}  with the following coefficients 
\begin{subequations}
\label{coefs in ev eq} 
\begin{equation}
\begin{array}{ll}
C(\visrat,  j,m)=&d(\visrat, j)^{-1}\left[c^\infty_{jm0}\sqrt{j(j+1)}\left(2j+1\right)\right.\\
&\left.+c^\infty_{jm2}\left(4j^3+6j^2-4j-3\right)\right]
\end{array}
\end{equation}
and
\begin{equation}
\label{Gamma}
\Gamma\left(\visrat, \sigma_0, j\right)=-E(j, \sigma_0)\frac{j(j+1)}{d(\visrat, j)}
\end{equation}
where 
\begin{equation}
d(\visrat, j) ={9\visrat^{-1}+\left(-5+3j^2+2j^3\right)}
\end{equation}
and
\begin{equation}
\label{Ej}
E(j, \sigma)=\left(j+2\right)\left(j-1\right)\left(j(j+1)+\sigma\right)\,.
\end{equation}
\end{subequations}
For a simple shear flow \refeq{shear flow}, solving Eq.\refeq{ev equation f} for  the stationary amplitudes in the tank-treading regime  gives
\begin{equation}
\label{amplitudes:1app}
f_{2\pm2}=\mp\frac{\im \chi 10\sqrt{\frac{2\pi}{15}}}{E(2, \sigma_0)\mp\im \chi\frac{(9+23\visrat)}{6}}
\end{equation}
and for the inclination angle
\begin{equation}
\label{angle:shear}
\phi_0=\frac{1}{2}\arctan\left[\frac{6 E(2, \sigma_0)}{\chi\left(9+23\visrat\right)}\right]\,.
\end{equation}
These expressions agree with Seifert's results  for a vesicle with no viscosity contrast ($\visrat=2)$ \cite{Seifert:1999}.
At long times, the excess area is stored in the $f_{2\pm2}$ modes only. Substituting  the tension \refeq{tension} in \refeq{Ej} leads to
\begin{equation}
\label{prolate}
E(2, \sigma_0)=\chi \frac{(23\visrat+9)}{6} \sqrt{-1+\frac{1920\pi}{ (23\visrat+9)^2\Delta}}.\,
\end{equation}
Thus the dependence of the mode amplitudes on the capillary number in Eq.\refeq{amplitudes:1app} cancels.
Expanding Eq.\ \refeq{angle:shear} for small values of the excess area $\Delta$
we obtain \refeq{angle for small Delta}.

The surface solenoidal velocity field $\bu^\pm_{jm1}$ satisfies identically the incompressibility condition \refeq{membrane incompressibility}.  The stress boundary condition 
\begin{equation}
\tau^\out_{jm1}-\left(\visrat-1\right)\tau^{in}_{jm1}=0
\end{equation}
 corresponds to a spherical viscous drop with no surface tension, i.e. the  $\bu^\pm_{jm1}$ flow field is unaffected by the interfacial stresses. Consequently,  for the   $j=1$  component of the shear flow, which corresponds to rigid body rotation, we have that  $c_{1m1}=c_{1m1}^\infty$, i.e. at leading order the particle rotates with the flow.

For the sake of completeness we mention the relation between our notation and the one used by Seifert\cite{Seifert:1999}
\begin{equation}
\begin{split}
c_{jm2}=&X_{jm}\,,\\
c_{jm0}=&\left(Y_{jm}+2X_{jm}\right)\left[j(j+1)\right]^{-\half}\,.
\end{split}
\end{equation}
\section{The isotropic tension}
\label{Area constraint}
The area constraint \refeq{delta to second order} serves to determine the isotropic part of the tension $\sigma_0$. We can split Eq.~\refeq{Gamma} into 
\begin{equation}
\Gamma(\visrat,\sigma_0,j)=\alpha(\visrat,j)+\sigma_0\beta(\visrat,j)\,,
\end{equation}
where
\begin{equation}
\alpha(\visrat,j)=-2a(j)\frac{[j(j+1)]^2}{d(\visrat,j)} \,,
\end{equation}
and
\begin{equation}
\beta(\visrat,j)=-2a(j)\frac{j(j+1)}{d(\visrat,j)}\,.
\end{equation}
The modes have to satisfy the area constraint $\dot \Delta=0$.
Multiplying the evolution equations \refeq{ev equation f} by $a(j)f^*_{jm}$ , summing up and solving for the tension leads to
\begin{widetext}
\begin{equation}
\label{tension}
\sigma_0=-\frac{\sum_{jm} a(j) \left[C(\visrat, j,m)f^*_{jm}+ \chi^{-1}  \alpha(\visrat,j) f_{jm}f^*_{jm}\right]}{\chi^{-1}\sum_{jm} a(j) \beta(\visrat,j) f_{jm}f^*_{jm}}\,.
\end{equation}
\end{widetext}
The complicated dependence of the tension on the shape modes makes the evolution equations  Eq.\ \refeq{ev equation f}  nonlinear. 


\bibliography{/thayerfs/home/petia_vlahovska/refs/refs}

\begin{thebibliography}{50}
\expandafter\ifx\csname natexlab\endcsname\relax\def\natexlab#1{#1}\fi
\expandafter\ifx\csname bibnamefont\endcsname\relax
  \def\bibnamefont#1{#1}\fi
\expandafter\ifx\csname bibfnamefont\endcsname\relax
  \def\bibfnamefont#1{#1}\fi
\expandafter\ifx\csname citenamefont\endcsname\relax
  \def\citenamefont#1{#1}\fi
\expandafter\ifx\csname url\endcsname\relax
  \def\url#1{\texttt{#1}}\fi
\expandafter\ifx\csname urlprefix\endcsname\relax\def\urlprefix{URL }\fi
\providecommand{\bibinfo}[2]{#2}
\providecommand{\eprint}[2][]{\url{#2}}

\bibitem[{\citenamefont{Rallison}(1984)}]{Rallison:1984}
\bibinfo{author}{\bibfnamefont{J.~M.} \bibnamefont{Rallison}},
  \bibinfo{journal}{Ann. Rev. Fluid Mech.} \textbf{\bibinfo{volume}{16}},
  \bibinfo{pages}{45} (\bibinfo{year}{1984}).

\bibitem[{\citenamefont{Stone}(1994)}]{Stone:1994}
\bibinfo{author}{\bibfnamefont{H.~A.} \bibnamefont{Stone}},
  \bibinfo{journal}{Ann. Rev. Fluid Mech.} \textbf{\bibinfo{volume}{26}},
  \bibinfo{pages}{65} (\bibinfo{year}{1994}).

\bibitem[{\citenamefont{Barthes-Biesel}(1991)}]{Barthes_Biesel:1991}
\bibinfo{author}{\bibfnamefont{D.}~\bibnamefont{Barthes-Biesel}},
  \bibinfo{journal}{Physica A} \textbf{\bibinfo{volume}{172}},
  \bibinfo{pages}{103} (\bibinfo{year}{1991}).

\bibitem[{\citenamefont{Canham}(1970)}]{Canham:1970}
\bibinfo{author}{\bibfnamefont{P.}~\bibnamefont{Canham}}, \bibinfo{journal}{J.
  Theor. Biol.} \textbf{\bibinfo{volume}{26}}, \bibinfo{pages}{61}
  (\bibinfo{year}{1970}).

\bibitem[{\citenamefont{Powers}(2005)}]{Tom:review}
\bibinfo{author}{\bibfnamefont{T.~R.} \bibnamefont{Powers}}, in
  \emph{\bibinfo{booktitle}{Handbook of Materials Modeling}}, edited by
  \bibinfo{editor}{\bibfnamefont{S.}~\bibnamefont{Yip}}
  (\bibinfo{publisher}{Springer}, \bibinfo{year}{2005}), pp.
  \bibinfo{pages}{2631--2643}.

\bibitem[{\citenamefont{Seifert}(1997)}]{Seifert:1997}
\bibinfo{author}{\bibfnamefont{U.}~\bibnamefont{Seifert}},
  \bibinfo{journal}{Advances in physics} \textbf{\bibinfo{volume}{46}},
  \bibinfo{pages}{13} (\bibinfo{year}{1997}).

\bibitem[{\citenamefont{de~Haas et~al.}(1997)\citenamefont{de~Haas, Blom,
  van~den Ende, Duits, and Mellema}}]{Haas-Blom-Ende-Duits-Mellema:1997}
\bibinfo{author}{\bibfnamefont{K.~H.} \bibnamefont{de~Haas}},
  \bibinfo{author}{\bibfnamefont{C.}~\bibnamefont{Blom}},
  \bibinfo{author}{\bibfnamefont{D.}~\bibnamefont{van~den Ende}},
  \bibinfo{author}{\bibfnamefont{M.~H.~G.} \bibnamefont{Duits}},
  \bibnamefont{and} \bibinfo{author}{\bibfnamefont{J.}~\bibnamefont{Mellema}},
  \bibinfo{journal}{Phys. Rev. E} \textbf{\bibinfo{volume}{56}},
  \bibinfo{pages}{7132} (\bibinfo{year}{1997}).

\bibitem[{\citenamefont{Kantsler and
  Steinberg}(2005)}]{Kantsler-Steinberg:2005}
\bibinfo{author}{\bibfnamefont{V.}~\bibnamefont{Kantsler}} \bibnamefont{and}
  \bibinfo{author}{\bibfnamefont{V.}~\bibnamefont{Steinberg}},
  \bibinfo{journal}{Phys. Rev. Lett.} \textbf{\bibinfo{volume}{95}},
  \bibinfo{pages}{258101} (\bibinfo{year}{2005}).

\bibitem[{\citenamefont{Kantsler and
  Steinberg}(2006)}]{Kantsler-Steinberg:2006}
\bibinfo{author}{\bibfnamefont{V.}~\bibnamefont{Kantsler}} \bibnamefont{and}
  \bibinfo{author}{\bibfnamefont{V.}~\bibnamefont{Steinberg}},
  \bibinfo{journal}{Phys. Rev. Lett.} \textbf{\bibinfo{volume}{96}},
  \bibinfo{pages}{036001} (\bibinfo{year}{2006}).

\bibitem[{\citenamefont{Mader et~al.}(2006)\citenamefont{Mader, Vitkova,
  Abkarian, Viallat, and Podgorski}}]{Mader:2006}
\bibinfo{author}{\bibfnamefont{M.-A.} \bibnamefont{Mader}},
  \bibinfo{author}{\bibfnamefont{V.}~\bibnamefont{Vitkova}},
  \bibinfo{author}{\bibfnamefont{M.}~\bibnamefont{Abkarian}},
  \bibinfo{author}{\bibfnamefont{A.}~\bibnamefont{Viallat}}, \bibnamefont{and}
  \bibinfo{author}{\bibfnamefont{T.}~\bibnamefont{Podgorski}},
  \bibinfo{journal}{Eur. Phys. J. E} \textbf{\bibinfo{volume}{19}},
  \bibinfo{pages}{389} (\bibinfo{year}{2006}).

\bibitem[{\citenamefont{Kraus et~al.}(1996)\citenamefont{Kraus, Wintz, Seifert,
  and Lipowsky}}]{Kraus-Seifert_etal:1996}
\bibinfo{author}{\bibfnamefont{M.}~\bibnamefont{Kraus}},
  \bibinfo{author}{\bibfnamefont{W.}~\bibnamefont{Wintz}},
  \bibinfo{author}{\bibfnamefont{U.}~\bibnamefont{Seifert}}, \bibnamefont{and}
  \bibinfo{author}{\bibfnamefont{R.}~\bibnamefont{Lipowsky}},
  \bibinfo{journal}{Phys.Rev.Lett.} \textbf{\bibinfo{volume}{77}},
  \bibinfo{pages}{3685} (\bibinfo{year}{1996}).

\bibitem[{\citenamefont{Beaucourt et~al.}(2004)\citenamefont{Beaucourt, Rioual,
  Seon, Biben, and Misbah}}]{Beaucourt_etal:2004}
\bibinfo{author}{\bibfnamefont{J.}~\bibnamefont{Beaucourt}},
  \bibinfo{author}{\bibfnamefont{F.}~\bibnamefont{Rioual}},
  \bibinfo{author}{\bibfnamefont{T.}~\bibnamefont{Seon}},
  \bibinfo{author}{\bibfnamefont{T.}~\bibnamefont{Biben}}, \bibnamefont{and}
  \bibinfo{author}{\bibfnamefont{C.}~\bibnamefont{Misbah}},
  \bibinfo{journal}{Phys. Rev. E} \textbf{\bibinfo{volume}{69}},
  \bibinfo{pages}{011906} (\bibinfo{year}{2004}).

\bibitem[{\citenamefont{Noguchi and Gompper}(2004)}]{Noguchi-Gompper:2004}
\bibinfo{author}{\bibfnamefont{H.}~\bibnamefont{Noguchi}} \bibnamefont{and}
  \bibinfo{author}{\bibfnamefont{G.}~\bibnamefont{Gompper}},
  \bibinfo{journal}{Phys. Rev. Lett.} \textbf{\bibinfo{volume}{93}},
  \bibinfo{pages}{258102} (\bibinfo{year}{2004}).

\bibitem[{\citenamefont{Seifert}(1999{\natexlab{a}})}]{Seifert:1999}
\bibinfo{author}{\bibfnamefont{U.}~\bibnamefont{Seifert}},
  \bibinfo{journal}{Eur. Phys. J. B} \textbf{\bibinfo{volume}{8}},
  \bibinfo{pages}{405} (\bibinfo{year}{1999}{\natexlab{a}}).

\bibitem[{\citenamefont{Olla}(2000)}]{Olla:2000}
\bibinfo{author}{\bibfnamefont{P.}~\bibnamefont{Olla}},
  \bibinfo{journal}{Physica A} \textbf{\bibinfo{volume}{278}},
  \bibinfo{pages}{87} (\bibinfo{year}{2000}).

\bibitem[{\citenamefont{Misbah}(2006)}]{Misbah:2006}
\bibinfo{author}{\bibfnamefont{C.}~\bibnamefont{Misbah}},
  \bibinfo{journal}{Phys. Rev. Lett.} \textbf{\bibinfo{volume}{96}},
  \bibinfo{pages}{028104} (\bibinfo{year}{2006}).

\bibitem[{\citenamefont{Keller and Skalak}(1982)}]{Keller-Skalak:1982}
\bibinfo{author}{\bibfnamefont{S.~R.} \bibnamefont{Keller}} \bibnamefont{and}
  \bibinfo{author}{\bibfnamefont{R.}~\bibnamefont{Skalak}},
  \bibinfo{journal}{J. Fluid Mech.} \textbf{\bibinfo{volume}{120}},
  \bibinfo{pages}{27} (\bibinfo{year}{1982}).

\bibitem[{\citenamefont{Sutera and Skalak}(1993)}]{Sutera--Skalak:1993}
\bibinfo{author}{\bibfnamefont{S.~P.} \bibnamefont{Sutera}} \bibnamefont{and}
  \bibinfo{author}{\bibfnamefont{R.}~\bibnamefont{Skalak}},
  \bibinfo{journal}{Annu. Rev. Fluid Mech.} \textbf{\bibinfo{volume}{25}},
  \bibinfo{pages}{1} (\bibinfo{year}{1993}).

\bibitem[{\citenamefont{Popel and Johnson}(2005)}]{Popel:2005}
\bibinfo{author}{\bibfnamefont{A.~S.} \bibnamefont{Popel}} \bibnamefont{and}
  \bibinfo{author}{\bibfnamefont{P.~C.} \bibnamefont{Johnson}},
  \bibinfo{journal}{Annu. Rev. Fluid Mech.} \textbf{\bibinfo{volume}{37}},
  \bibinfo{pages}{43} (\bibinfo{year}{2005}).

\bibitem[{\citenamefont{Orsello et~al.}(2001)\citenamefont{Orsello,
  Lauffenburger, and Hammer}}]{Orsello-Hammer:2001}
\bibinfo{author}{\bibfnamefont{C.~E.} \bibnamefont{Orsello}},
  \bibinfo{author}{\bibfnamefont{D.~A.} \bibnamefont{Lauffenburger}},
  \bibnamefont{and} \bibinfo{author}{\bibfnamefont{D.~A.}
  \bibnamefont{Hammer}}, \bibinfo{journal}{Trends in Biotechnology}
  \textbf{\bibinfo{volume}{19}}, \bibinfo{pages}{310} (\bibinfo{year}{2001}).

\bibitem[{\citenamefont{Lorz et~al.}(2000)\citenamefont{Lorz, Simson, Nardi,
  and Sakmann}}]{Lorz-Simson-Nardi-Sakmann:2000}
\bibinfo{author}{\bibfnamefont{B.}~\bibnamefont{Lorz}},
  \bibinfo{author}{\bibfnamefont{R.}~\bibnamefont{Simson}},
  \bibinfo{author}{\bibfnamefont{J.}~\bibnamefont{Nardi}}, \bibnamefont{and}
  \bibinfo{author}{\bibfnamefont{E.}~\bibnamefont{Sakmann}},
  \bibinfo{journal}{Europhys. Lett.} \textbf{\bibinfo{volume}{51}},
  \bibinfo{pages}{468} (\bibinfo{year}{2000}).

\bibitem[{\citenamefont{Abkarian et~al.}(2002)\citenamefont{Abkarian, Lartigue,
  and Viallat}}]{Abkarian:2002}
\bibinfo{author}{\bibfnamefont{M.}~\bibnamefont{Abkarian}},
  \bibinfo{author}{\bibfnamefont{C.}~\bibnamefont{Lartigue}}, \bibnamefont{and}
  \bibinfo{author}{\bibfnamefont{A.}~\bibnamefont{Viallat}},
  \bibinfo{journal}{Phys. Rev. Lett.} \textbf{\bibinfo{volume}{88}},
  \bibinfo{pages}{068103} (\bibinfo{year}{2002}).

\bibitem[{\citenamefont{Abkarian and Viallat}(2005)}]{Abkarian:2005}
\bibinfo{author}{\bibfnamefont{M.}~\bibnamefont{Abkarian}} \bibnamefont{and}
  \bibinfo{author}{\bibfnamefont{A.}~\bibnamefont{Viallat}},
  \bibinfo{journal}{Biophys. J.} \textbf{\bibinfo{volume}{89}},
  \bibinfo{pages}{1055} (\bibinfo{year}{2005}).

\bibitem[{\citenamefont{Cantat and Misbah}(1999)}]{Cantat-Misbah:1999}
\bibinfo{author}{\bibfnamefont{I.}~\bibnamefont{Cantat}} \bibnamefont{and}
  \bibinfo{author}{\bibfnamefont{C.}~\bibnamefont{Misbah}},
  \bibinfo{journal}{Phys. Rev. Lett.} \textbf{\bibinfo{volume}{83}},
  \bibinfo{pages}{235} (\bibinfo{year}{1999}).

\bibitem[{\citenamefont{Sukumaran and Seifert}(2001)}]{Sukumaran-Seifert:2001}
\bibinfo{author}{\bibfnamefont{S.}~\bibnamefont{Sukumaran}} \bibnamefont{and}
  \bibinfo{author}{\bibfnamefont{U.}~\bibnamefont{Seifert}},
  \bibinfo{journal}{Phys. Rev. E} \textbf{\bibinfo{volume}{64}},
  \bibinfo{pages}{011916} (\bibinfo{year}{2001}).

\bibitem[{\citenamefont{Seifert}(1999{\natexlab{b}})}]{Seifert:1999b}
\bibinfo{author}{\bibfnamefont{U.}~\bibnamefont{Seifert}},
  \bibinfo{journal}{Phys. Rev. Lett.} \textbf{\bibinfo{volume}{83}},
  \bibinfo{pages}{876} (\bibinfo{year}{1999}{\natexlab{b}}).

\bibitem[{\citenamefont{Leal}(1992)}]{Leal:1992}
\bibinfo{author}{\bibfnamefont{L.~G.} \bibnamefont{Leal}},
  \emph{\bibinfo{title}{Laminar Flow and Convective Transport Processes}}
  (\bibinfo{publisher}{Butterworth-Heinemann}, \bibinfo{address}{Boston},
  \bibinfo{year}{1992}).

\bibitem[{\citenamefont{Seifert}(1995)}]{Seifert:1995}
\bibinfo{author}{\bibfnamefont{U.}~\bibnamefont{Seifert}},
  \bibinfo{journal}{Z.\ Phys.\ B} \textbf{\bibinfo{volume}{97}},
  \bibinfo{pages}{299} (\bibinfo{year}{1995}).

\bibitem[{\citenamefont{Vlahovska et~al.}(2005)\citenamefont{Vlahovska,
  B{\l}awzdziewicz, and Loewenberg}}]{Vlahovska:2005}
\bibinfo{author}{\bibfnamefont{P.}~\bibnamefont{Vlahovska}},
  \bibinfo{author}{\bibfnamefont{J.}~\bibnamefont{B{\l}awzdziewicz}},
  \bibnamefont{and}
  \bibinfo{author}{\bibfnamefont{M.}~\bibnamefont{Loewenberg}},
  \bibinfo{journal}{Phys.\ Fluids} \textbf{\bibinfo{volume}{17}},
  \bibinfo{pages}{Art. No.103103} (\bibinfo{year}{2005}).

\bibitem[{\citenamefont{Vlahovska}(2003)}]{Vlahovska:2003}
\bibinfo{author}{\bibfnamefont{P.}~\bibnamefont{Vlahovska}}, Ph.D. thesis,
  \bibinfo{school}{Yale University} (\bibinfo{year}{2003}), \bibinfo{note}{pdf
  file available by email: petia@aya.yale.edu}.

\bibitem[{\citenamefont{Barth{\`e}s-Biesel and
  Acrivos}(1973)}]{Barthes_Biesel-Acrivos:1973}
\bibinfo{author}{\bibfnamefont{D.}~\bibnamefont{Barth{\`e}s-Biesel}}
  \bibnamefont{and} \bibinfo{author}{\bibfnamefont{A.}~\bibnamefont{Acrivos}},
  \bibinfo{journal}{J. Fluid Mech.} \textbf{\bibinfo{volume}{61}},
  \bibinfo{pages}{1} (\bibinfo{year}{1973}).

\bibitem[{\citenamefont{Barthes-Biesel}(1980)}]{Barthes_Biesel:1980}
\bibinfo{author}{\bibfnamefont{D.}~\bibnamefont{Barthes-Biesel}},
  \bibinfo{journal}{J. Fluid Mech.} \textbf{\bibinfo{volume}{100}},
  \bibinfo{pages}{831} (\bibinfo{year}{1980}).

\bibitem[{\citenamefont{Biben and Misbah}(2003)}]{Biben-Misbah:2003}
\bibinfo{author}{\bibfnamefont{T.}~\bibnamefont{Biben}} \bibnamefont{and}
  \bibinfo{author}{\bibfnamefont{C.}~\bibnamefont{Misbah}},
  \bibinfo{journal}{Phys. Rev. E} \textbf{\bibinfo{volume}{67}},
  \bibinfo{pages}{031908} (\bibinfo{year}{2003}).

\bibitem[{\citenamefont{Kim and Karrila}(1991)}]{Kim-Karrila:1991}
\bibinfo{author}{\bibfnamefont{S.}~\bibnamefont{Kim}} \bibnamefont{and}
  \bibinfo{author}{\bibfnamefont{S.~J.} \bibnamefont{Karrila}},
  \emph{\bibinfo{title}{Microhydrodynamics: Principles and Selected
  Applications}} (\bibinfo{publisher}{Butterworth-Heinemann},
  \bibinfo{address}{London}, \bibinfo{year}{1991}).

\bibitem[{\citenamefont{Pozrikidis}(1992)}]{Pozrikidis:BIM}
\bibinfo{author}{\bibfnamefont{C.}~\bibnamefont{Pozrikidis}},
  \emph{\bibinfo{title}{Boundary Integral and Singularity Methods for
  Linearized Viscous Flow}} (\bibinfo{publisher}{Cambridge University Press},
  \bibinfo{year}{1992}).

\bibitem[{\citenamefont{B{\l}awzdziewicz
  et~al.}(2000)\citenamefont{B{\l}awzdziewicz, Vlahovska, and
  Loewenberg}}]{Blawzdziewicz-Vlahovska-Loewenberg:2000}
\bibinfo{author}{\bibfnamefont{J.}~\bibnamefont{B{\l}awzdziewicz}},
  \bibinfo{author}{\bibfnamefont{P.}~\bibnamefont{Vlahovska}},
  \bibnamefont{and}
  \bibinfo{author}{\bibfnamefont{M.}~\bibnamefont{Loewenberg}},
  \bibinfo{journal}{Physica A} \textbf{\bibinfo{volume}{276}},
  \bibinfo{pages}{50} (\bibinfo{year}{2000}).

\bibitem[{\citenamefont{Leal}(1980)}]{Leal:1980}
\bibinfo{author}{\bibfnamefont{L.~G.} \bibnamefont{Leal}},
  \bibinfo{journal}{Ann. Rev. Fluid Mech} \textbf{\bibinfo{volume}{12}},
  \bibinfo{pages}{435} (\bibinfo{year}{1980}).

\bibitem[{\citenamefont{Smart and Leighton}(1990)}]{Smart-Leighton:1990}
\bibinfo{author}{\bibfnamefont{J.~R.} \bibnamefont{Smart}} \bibnamefont{and}
  \bibinfo{author}{\bibfnamefont{D.~T.} \bibnamefont{Leighton}},
  \bibinfo{journal}{Phys. Fluids A} \textbf{\bibinfo{volume}{3}},
  \bibinfo{pages}{21} (\bibinfo{year}{1990}).

\bibitem[{\citenamefont{Olla}(1997)}]{Olla:1997a}
\bibinfo{author}{\bibfnamefont{P.}~\bibnamefont{Olla}},
  \bibinfo{journal}{J.Phys. II France} \textbf{\bibinfo{volume}{7}},
  \bibinfo{pages}{1533} (\bibinfo{year}{1997}).

\bibitem[{\citenamefont{Pozrikidis}(1995)}]{Pozrikidis:1995}
\bibinfo{author}{\bibfnamefont{P.}~\bibnamefont{Pozrikidis}},
  \bibinfo{journal}{J. Fluid Mech.} \textbf{\bibinfo{volume}{297}},
  \bibinfo{pages}{123} (\bibinfo{year}{1995}).

\bibitem[{\citenamefont{Eggleton and Popel}(1998)}]{Eggleton-Popel:1998}
\bibinfo{author}{\bibfnamefont{C.~D.} \bibnamefont{Eggleton}} \bibnamefont{and}
  \bibinfo{author}{\bibfnamefont{A.~S.} \bibnamefont{Popel}},
  \bibinfo{journal}{Phys. Fluids} \textbf{\bibinfo{volume}{10}},
  \bibinfo{pages}{1834} (\bibinfo{year}{1998}).

\bibitem[{\citenamefont{Lac et~al.}(2004)\citenamefont{Lac, Barthes-Biesel,
  Pelekasis, and Tsamopolous}}]{Lac-Barthes_Biesel:2004}
\bibinfo{author}{\bibfnamefont{E.}~\bibnamefont{Lac}},
  \bibinfo{author}{\bibfnamefont{D.}~\bibnamefont{Barthes-Biesel}},
  \bibinfo{author}{\bibfnamefont{N.}~\bibnamefont{Pelekasis}},
  \bibnamefont{and}
  \bibinfo{author}{\bibfnamefont{J.}~\bibnamefont{Tsamopolous}},
  \bibinfo{journal}{J. Fluid Mech.} \textbf{\bibinfo{volume}{516}},
  \bibinfo{pages}{303} (\bibinfo{year}{2004}).

\bibitem[{\citenamefont{Kwak~S}(2001)}]{Kwak-Pozrikidis:2001}
\bibinfo{author}{\bibfnamefont{P.~C.} \bibnamefont{Kwak~S}},
  \bibinfo{journal}{Phys. Fluids} \textbf{\bibinfo{volume}{13}},
  \bibinfo{pages}{1234} (\bibinfo{year}{2001}).

\bibitem[{\citenamefont{Pozrikidis}(2001)}]{Pozrikidis:2001}
\bibinfo{author}{\bibfnamefont{C.}~\bibnamefont{Pozrikidis}},
  \bibinfo{journal}{J. Fluid Mech.} \textbf{\bibinfo{volume}{440}},
  \bibinfo{pages}{269} (\bibinfo{year}{2001}).

\bibitem[{\citenamefont{Noguchi and Gompper}(2005)}]{Noguchi-Gompper:2005}
\bibinfo{author}{\bibfnamefont{H.}~\bibnamefont{Noguchi}} \bibnamefont{and}
  \bibinfo{author}{\bibfnamefont{G.}~\bibnamefont{Gompper}},
  \bibinfo{journal}{Phys. Rev. E} \textbf{\bibinfo{volume}{72}},
  \bibinfo{pages}{011901} (\bibinfo{year}{2005}).

\bibitem[{\citenamefont{Cristini et~al.}(2001)\citenamefont{Cristini,
  B{\l}awzdziewicz, and Loewenberg}}]{Cristini-Blawzdziewicz-Loewenberg:2001d}
\bibinfo{author}{\bibfnamefont{V.}~\bibnamefont{Cristini}},
  \bibinfo{author}{\bibfnamefont{J.}~\bibnamefont{B{\l}awzdziewicz}},
  \bibnamefont{and}
  \bibinfo{author}{\bibfnamefont{M.}~\bibnamefont{Loewenberg}},
  \bibinfo{journal}{J.\ Comput.\ Phys.} \textbf{\bibinfo{volume}{168}},
  \bibinfo{pages}{445} (\bibinfo{year}{2001}).

\bibitem[{\citenamefont{Jones}(1985)}]{Jones:1985}
\bibinfo{author}{\bibfnamefont{M.~N.} \bibnamefont{Jones}},
  \emph{\bibinfo{title}{Spherical Harmonics and Tensors for Classical Field
  Theory}} (\bibinfo{publisher}{Wiley}, \bibinfo{address}{New York},
  \bibinfo{year}{1985}).

\bibitem[{\citenamefont{Varshalovich et~al.}(1988)\citenamefont{Varshalovich,
  Moskalev, and Kheronskii}}]{Varshalovich:1988}
\bibinfo{author}{\bibfnamefont{D.~A.} \bibnamefont{Varshalovich}},
  \bibinfo{author}{\bibfnamefont{A.~N.} \bibnamefont{Moskalev}},
  \bibnamefont{and} \bibinfo{author}{\bibfnamefont{V.~K.}
  \bibnamefont{Kheronskii}}, \emph{\bibinfo{title}{Quantum Theory of Angular
  Momentum}} (\bibinfo{publisher}{World Scientfic},
  \bibinfo{address}{Singapore}, \bibinfo{year}{1988}).

\bibitem[{\citenamefont{B{\l}awzdziewicz
  et~al.}(1999)\citenamefont{B{\l}awzdziewicz, Wajnryb, and
  Loewenberg}}]{Blawzdziewicz-Wajnryb-Loewenberg:1999}
\bibinfo{author}{\bibfnamefont{J.}~\bibnamefont{B{\l}awzdziewicz}},
  \bibinfo{author}{\bibfnamefont{E.}~\bibnamefont{Wajnryb}}, \bibnamefont{and}
  \bibinfo{author}{\bibfnamefont{M.}~\bibnamefont{Loewenberg}},
  \bibinfo{journal}{J. Fluid Mech.} \textbf{\bibinfo{volume}{395}},
  \bibinfo{pages}{29} (\bibinfo{year}{1999}).

\bibitem[{\citenamefont{Cichocki et~al.}(1988)\citenamefont{Cichocki,
  Felderhof, and Schmitz}}]{Cichocki-Felderhof-Schmitz:1988}
\bibinfo{author}{\bibfnamefont{B.}~\bibnamefont{Cichocki}},
  \bibinfo{author}{\bibfnamefont{B.~U.} \bibnamefont{Felderhof}},
  \bibnamefont{and} \bibinfo{author}{\bibfnamefont{R.}~\bibnamefont{Schmitz}},
  \bibinfo{journal}{PhysicoChem. Hyd.} \textbf{\bibinfo{volume}{10}},
  \bibinfo{pages}{383} (\bibinfo{year}{1988}).

\end{thebibliography}
\end{document}